\documentclass{conm-p-l}
\usepackage{amsthm,amsfonts,latexsym,amssymb
}
\def\a{\alpha}

\def\c{\gamma}

\def\e{\varepsilon}
\def\s{\sigma}

\def\D{\Delta}

\def\p{\partial}
\def\x{{\hat{x}}}

\def\R{\mathbb R}

\newcommand{\be}{\begin{equation}}
\newcommand{\ee}{\end{equation}}
\newcommand{\bea}{\begin{eqnarray}}
\newcommand{\eea}{\end{eqnarray}}
\newcommand{\beax}{\begin{eqnarray*}}
\newcommand{\eeax}{\end{eqnarray*}}

\newcommand{\Tr}{\mbox{\rm Tr}}
\newcommand{\mfr}[2]{{\textstyle\frac{#1}{#2}}}
\newcommand{\V}{V^{\rm TF}}

\newtheorem{theorem}{Theorem}[section]
\newtheorem{lemma}[theorem]{Lemma}

\theoremstyle{definition}

\theoremstyle{remark}

\numberwithin{equation}{section}

\begin{document}

\title{New coherent states and a new proof of the Scott correction}

\author{Jan Philip Solovej}
\address{Department of Mathematics, University of Copenhagen,
Universitetsparken 5, DK-2100 Copenhagen, Denmark}
\email{solovej@math.ku.dk}
\thanks{ \copyright 2002 \ by the authors. This article may be
      reproduced in its entirety for non-commercial purposes.}

\author{Wolfgang L Spitzer}
\address{Department of Mathematics, University of California, One Shields Avenue,
Davis 95616-8366, USA}
\email{spitzer@math.ucdavis.edu}

\date{August 31, 2002}



\begin{abstract}
We introduce new coherent states and use them to prove semi-classical estimates
for Schr\"odinger operators with regular potentials. This can be further applied 
to the Thomas-Fermi potential yielding a new proof of the Scott correction for molecules.
\end{abstract}

\maketitle

\section{Introduction}

In this paper we review a novel proof of the Scott correction for neutral mol\-ecules. So
suppose, we have $M$ nuclei of positive charges $Z=(Z_1,\ldots, Z_M)\in\R_+^M$ located at
positions $R=(R_1,\ldots,R_M)\in\R^{3N}$. We choose the charge of an electron equal to $-1$,
so that neutrality is expressed as $|Z|=\sum_{j=1}^M Z_j = N$, where $N$ is the number of
electrons. Further, we use atomic units where $\hbar^2=m$.

The interaction of a single electron with all the nuclei is equal to
\begin{equation}\label{V}
   V(Z,R,x) = \sum_{j=1}^M \frac{Z_j}{|x-R_j|} .
\end{equation}

We now write the molecular non-relativistic Schr\"odinger operator in the form
\beax\lefteqn{H(Z,R)=H(Z_1,\ldots,Z_M;R_1,\ldots, R_M)
}
\\
&=&\sum_{i=1}^N\left(-\mfr{1}{2}\D_i - V(Z,R,x_i)\right)
    +\sum_{1\le i<j\le N}\frac{1}{|x_i-x_j|}.
\eeax
The operator $H(Z,R)$ acts on the space $\bigwedge_{i=1}^N L^2(\R^3\times\{-1,1\})$,
where $\pm1$ refers to the spin variables. We are interested in the ground state energy,
\begin{equation} E(Z,R) = \inf\mbox{spec} H(Z,R) ,
\end{equation}
and in particular, in an asymptotic expansion for large charges. Let us state the
main theorem in this paper.

\begin{theorem}[Scott correction] \label{main theorem}
Let $Z=|Z|(z_1,\ldots,z_M)$, where $z_1,\ldots,z_M$ $>0$ and $R=|Z|^{-1/3}(r_1,\ldots,r_M)$, 
with $|r_i - r_j|>r_0>0$, for all $i\ne j$. Then,
\begin{equation}
  E(Z,R) = E^{\mbox\tiny TF} (Z,R) + \mfr{1}{2} \sum_{1\le j\le M} Z_j^2 +
  {\mathcal O} (|Z|^{2-1/30}),
\end{equation}
as $|Z|\to\infty$, where the error term ${\mathcal O} (|Z|^{2-1/30})$
besides $|Z|$ depends only on $z_1,\ldots,z_M$, and $r_0$.
\end{theorem}

The leading Thomas-Fermi (TF) term, which is of the order $|Z|^{7/3}$ was first
rigorously derived in the seminal work by Lieb and Simon~\cite{Lieb-Simon} using the
Dirichlet-Neumann bracketing method. 

The Scott correction, i.e., the term $\mfr{1}{2} \sum_{1\le j\le M} Z_j^2$ was
proven by Hughes~\cite{Hughes} (a lower bound), and by Siedentop and
Weikard~\cite{Siedentop-Weikard} (both bounds) in the case of atoms. The atomic case is simpler
since in TF theory atoms are spherically symmetric. Bach~\cite{Bach} proved the Scott correction
for ions. Finally, Ivrii and Sigal~\cite{Ivrii-Sigal} accomplished a proof of the Scott correction
for molecules, which was recently extended to matter by Balodis Matesanz~\cite{Matesanz}. 
Here, we present another proof for molecules.

It was later shown by Lieb~\cite{Lieb1} (and independently by Thirring~\cite{Thirring}) how 
coherent states can be used to give a simple proof of the leading TF term with good upper and 
lower bounds; see also a recent improvement by Balodis Matesanz and Solovej~\cite{Matesanz-Solovej}.
We want to stress that in order to prove
an asymptotic expansion for $E(Z,R)$ capturing the Scott term one basically needs to prove a
local trace formula for regular potentials (see Theorem~\ref{corollary} with $n=3$) 
up to the order $h^{-2+\e}$ where $\e$ is any positive number. We accomplish $\e=1/5$. 

A quick explanation for the $Z^2$-correction goes as follows. Whereas the leading TF term
comes from the bulk of electrons, the correction comes {\it only} from electrons
close to the nuclei where the Coulomb attraction is unscreened by the
presence of the other electrons. From the exact solution of the hydrogen atom one may extract
the Scott correction (see~\cite{Lieb1}). Notice, that the Scott correction for molecules is 
just the sum of the corresponding atomic corrections. This is not the case for the leading term.

This review is organized as follows. In Section 2 we recall the main analytic tools and state 
the main properties of the TF potential. We introduce the new coherent states in Section
3. In Section 4 we sketch the proof of the semi-classical estimates on the sum of negative 
eigenvalues for regular and the TF potential. In the last Section we present the proof of
the main Theorem~\ref{main theorem}. For more details we refer to our paper~\cite{SS}.

\section{Preliminaries}

\subsection{Some Inequalities}

Here we collect the main inequalities which we need in this paper. Various constants are typically
denoted by the same letter $C$, and in all cases sharp constants do not play a role.

Let $p\ge1$, then a complex-valued function $f$ (and only those will be considered here)
is said to be in $L^p(\R^n)$ if the norm
$\|f\|_p := \left(\int |f(x)|^p \,dx\right)^{1/p}$ is finite. For any $1\le p\le t\le
q\le\infty$ we have the inclusion $L^p\cap L^q\subset L^t$, since by H\"older's inequality
$\| f\|_t \le \|f\|_p^\lambda \|f\|_q^{1-\lambda}$ with $\lambda p^{-1}+(1-\lambda) q^{-1}
=t^{-1}$.

We call $\c$ a density matrix on $L^2(\R^n)$ if it is a trace class
operator on $L^2(\R^n)$ satisfying the operator inequality ${\bf
  0}\le\c\le {\bf 1}$. The density of a density matrix $\gamma$ is the
$L^1$ function $\rho_\gamma$ such that
$\Tr(\gamma\theta)=\int\rho_\gamma(x)\theta(x)dx$ for all $\theta\in
C_0^\infty(\R^n)$ considered as a multiplication operator.

If $\psi\in\bigotimes_{i=1}^N L^2(\R^3\times\{-1,1\})$ is an $N$-body wave-function, then
its one-particle density, $\rho_\psi$, is defined by
$$
\rho_\psi(x) = \sum_{i=1}^N \sum_{s_1=\pm1}\cdots\sum_{s_N=\pm1}\int
|\psi(x_1,s_1;\ldots;x_N,s_N)|^2\,\delta(x_i-x)\,dx_1\cdots x_N.
$$
The next inequality we recall is crucial to most of our estimates.

\begin{theorem}[Lieb-Thirring inequality] \label{Lieb-Thirring}
\begin{description}
\item[One-body case] Let $\c$ be a \\ density operator on $L^2(\R^n)$,
then we have the Lieb-Thirring (LT) inequality
\begin{equation}\label{LTdensity}
   \Tr\left[-\mfr{1}{2}\Delta \gamma\right]\geq K_n\int\rho_\gamma^{1+2/n},
\end{equation}
where $K_n$ is some positive constant. Equivalently, let $V\in L^{1+n/2}(\mathbb R^n)$ 
and $\c$ a density operator, then \begin{equation}\label{LT} \Tr
[(-\mfr{1}{2}\D + V)\c] \ge -L_n \int |V_-|^{1+n/2}, \end{equation} where $x_- :=
\min\{x,0\}$, and $L_n$ some positive constant.

\item[Many-body case] Let $\psi\in \bigwedge_{i=1}^N L^2(\R^{3}\times\{-1,1\})$. Then,
\begin{equation}\label{eq:LTmbcase}
  \left\langle\psi,\sum_{i=1}^N-\mfr{1}{2}\Delta_i\psi\right\rangle\geq
  2^{-2/3} K_3\int\rho_\psi^{5/3} .
\end{equation}
\end{description}
\end{theorem}

The original proofs of these inequalities can be found in
\cite{Lieb-Thirring}. From the min-max principle it is clear
that the right hand side of (\ref{LT}) is in fact a lower bound on the sum
of the negative eigenvalues of the operator $-\frac{1}{2}\Delta + V$.

We shall use the following standard notation for the Coulomb energy:
$$D(f)=D(f,f)=\frac{1}{2}\int\!\!\int \bar{f}(x)|x-y|^{-1} f(y)\, dx dy .
$$
It is not difficult to see (by Fourier transformation) that $\|f\| := D(f)^{1/2}$ is a norm.
\begin{theorem}[Hardy-Littlewood-Sobolev inequality] There exists a constant $C$ such that
\begin{equation} \label{Hardy-Littlewood-Sobolev}
    D(f)\le C\,\| f \|_{6/5}^2.
\end{equation}
\end{theorem}
The sharp constant $C$ has been found by Lieb \cite{Lieb:sob}, see also \cite{Lieb-Loss}.

Finally, we state the two inequalities which we shall need to estimate the many-body ground state
energy, $E(Z,R)$, by an energy of an effective one-particle quantum system.
The first one is the electrostatic inequality providing us with a lower bound.
This inequality is due to Lieb \cite{Lieb3}, and was improved in \cite{Lieb-Oxford}.

\begin{theorem}[Lieb-Oxford inequality] Let $\psi\in L^2(\R^{3N})$ be normalized, and
$\rho_\psi$ its one-electron density. Then,
\begin{equation}\label{Lieb-Oxford} \left\langle \psi,\sum_{1\le i<j\le N} |x_i -x_j|^{-1}
   \psi\right\rangle \ge D(\rho_\psi) - C \int \rho_\psi^{4/3} .
\end{equation}
\end{theorem}

An upper bound to $E(Z,R)$ is furnished by a variational principle for
fermionic systems. This is also due to Lieb \cite{Lieb2}.

\begin{theorem}[Lieb's Variational Principle] \label{Lieb's Variational Principle}
Let $\c$ be a density operator on $L^2(\R^3)$ satisfying
$2\Tr\c= 2\int \rho_\c(x)\, dx \leq Z$ (i.e., less than or equal to the number of
electrons) with the kernel $\rho_\c(x)=\c(x,x)$. Then,
\begin{equation} \label{variational principle}
E(Z,R)\le 2\Tr\Big[\left(-\mfr{1}{2}\D - V(Z,R,x)\right)\c\Big] + D(2\rho_\gamma) .
\end{equation}
\end{theorem}

The factors 2 above are due to the spin degeneracy. The ground state wave-function
carries a spin and is really a function on $L^2(\R^{3Z}; \mathbb C^{2^Z})$, but only its
spatial dependency is of interest here.

\subsection{Thomas-Fermi Theory}
Here we quickly state the properties about TF theory which are needed for our proof.
The original proofs can be found in \cite{Lieb-Simon} and \cite{Lieb1}.

Consider ${\mathbf z}=(z_1,\ldots,z_M)\in\R_+^M$ and
${\mathbf r}=(r_1,\ldots,r_M)\in\R^{3M}$.  Let
$0\le\rho\in L^{5/3}(\mathbb R^3)\cap L^1(\mathbb R^3)$ then the
TF energy functional, ${\mathcal E}^{{\rm TF}}$ (omitting the nuclei-nuclei
interactions), is defined as
\begin{equation}\label{def:TF}
{\mathcal E}^{{\rm TF}}(\rho) =
\mfr{3}{10}(3\pi^2)^{2/3}\int \rho(x)^{5/3}\, dx - \int
V({\mathbf z},{\mathbf r},x)\rho(x)\, dx+ D(\rho) ,
\end{equation}
where $V$ is as in (\ref{V}).

\begin{theorem}[Thomas-Fermi minimizer]
For all ${\mathbf z}=(z_1,\ldots,z_M)\in \R_+^M$ and
${\mathbf r}=(r_1,\ldots,r_M)\in\R^{3M}$ there exists a unique non-negative
$\rho^{{\rm TF}}({\mathbf z},{\mathbf r},x)$
such that $\int \rho^{{\rm TF}}({\mathbf z},{\mathbf r},x)\,
dx=\sum_{k=1}^Mz_k$ and
$$
{\mathcal E}^{{\rm TF}}(\rho^{{\rm TF}}) = \inf
\left\{{\mathcal E}^{{\rm TF}} (\rho)\ :\ 0\le\rho\in L^{5/3}(\R^3)\cap L^{1}(\R^3)\right\} .
$$
We shall denote by
$E^{{\rm TF}}({\mathbf z},{\mathbf r}) := {\mathcal E}^{{\rm TF}} (\rho^{{\rm TF}})$
the TF-energy. Moreover, let
\begin{equation}\label{eq:tfpotgeneral}
 V^{{\rm TF}}({\mathbf z},{\mathbf r},x) := V({\mathbf
    z},{\mathbf r},x)
  -\rho^{{\rm TF}} * |x|^{-1}.
\end{equation}
be the TF-potential, then $V^{{\rm TF}}>0$ and $\rho^{{\rm TF}}>0$,
and $\rho^{{\rm TF}}$ is the unique solution in
$L^{5/3}(\R^3)\cap L^1(\R^3)$ to the TF-equation:
\begin{equation}\label{eq:tfeqgeneral}
   V^{{\rm TF}}({\mathbf z},{\mathbf r},x) =
   \mfr{1}{2}(3\pi^2)^{2/3}
   \rho^{{\rm TF}}({\mathbf z},{\mathbf r},x)^{2/3}.
\end{equation}
\end{theorem}

Very crucial for a semi-classical approach is the {\it scaling}
behavior of the TF-potential.  It says that for any positive parameter
$h$

\begin{eqnarray}\label{scaling}
  V^{{\rm TF}}({\mathbf z},{\mathbf r},x) &=&
  h^{-4} V^{{\rm TF}}(h^{3}{\mathbf z},h^{-1}{\mathbf r},h^{-1}x),
  \\
  \rho^{{\rm TF}}({\mathbf z},{\mathbf r},x) &=& h^{-6}\rho^{{\rm
      TF}}(h^{3}{\mathbf z},h^{-1}{\mathbf r},h^{-1}x) ,
  \\
  E^{\rm TF}({\mathbf z},{\mathbf r})&=&
  h^{-7}E^{\rm TF}(h^{3}{\mathbf z},h^{-1}{\mathbf r}).
\end{eqnarray}
By $h^{-1}{\mathbf r}$ (and likewise for $h^{3}{\mathbf z}$) we mean that each coordinate is 
scaled by $h^{-1}$. Notice that the Coulomb-potential, $V$, has the claimed scaling behavior. The
rest follows from the uniqueness of the solution of the TF-energy functional.

We shall now state the crucial estimates that we need about the TF potential. Let
\begin{equation}\label{ddefinition}
  d(x)=\min\{|x-r_k|\ |\ k=1,\ldots,M\} ,
\end{equation}
and
\begin{equation}\label{fdefinition}
  f(x)=\min\{d(x)^{-1/2}, d(x)^{-2}\}.
\end{equation}
For each $k=1,\ldots,M$ we define the function
\begin{equation}\label{eq:Wdefinition}
  W_k({\mathbf z},{\mathbf r},x)=V^{\rm TF}({\mathbf z},{\mathbf r},x)
  -z_k|x-r_k|^{-1}.
\end{equation}
The function $W_k$ can be continuously extended to $x=r_k$. We have the following estimate for the 
TF potential.

\begin{theorem}[Estimate on TF potential]\label{thm:tfestimate}
Let ${\mathbf z}=(z_1,\ldots,z_M)\in \R_+^M$ and
${\mathbf r}=(r_1,\ldots,r_M)\in \R^{3M}$.
For all multi-indices $\alpha$ and all $x$ with
$d(x)\ne0$ we have
\begin{equation}\label{eq:tfdf}
\left|\partial^\alpha_x\V({\mathbf z},{\mathbf r},x)\right|\leq C_\alpha f(x)^2
d(x)^{-|\alpha|},
\end{equation}
where $C_\alpha>0$ is a constant which depends on $\alpha$, $z_1,\ldots,z_M$,
and $M$.

Moreover, for $|x-r_k|<r_{\min}/2$, where
$r_{\min}=\min_{k\ne\ell}|r_k-r_\ell|$ we have
\begin{equation}\label{eq:westimate}
 0\leq W_k({\mathbf z},{\mathbf r},x)
  \leq Cr_{\min}^{-1}+C,
\end{equation}
where the constant $C>0$ here depends on $z_1,\ldots,z_M$,
and $M$.
\end{theorem}

The relation of TF theory to semi-classical analysis is that
the semi-classical density of a gas of non-interacting electrons moving
in the TF potential is simply the TF density.
More precisely, the semi-classical approximation to
the density of the projection onto the eigenspace corresponding to
negative eigenvalues of the Hamiltonian $-\frac{1}{2}\Delta-\V$ is
$$
2\int_{\frac{1}{2}p^2-\V({\mathbf z},{\mathbf r},x)\le0}1
\frac{dp}{(2\pi)^3} = 2^{3/2} (3\pi^2)^{-1} (\V)^{3/2}({\mathbf
  z},{\mathbf r},x)= \rho^{\rm TF}({\mathbf z},{\mathbf r},x).
$$
Here, the factor two on the very left is due to the spin degeneracy.
Similarly, the semi-classical approximation to the energy of the gas,
i.e., to the sum of the negative eigenvalues of
$-\frac{1}{2}\Delta-\V$ is
\bea\label{eq:sc=tf}
2\int \left(\frac{1}{2}p^2-\V({\mathbf z},{\mathbf r},x)\right)_-
\frac{dpdx}{(2\pi)^3}&=& -{\frac {4\sqrt {2}}{15{\pi }^{2}}}\int
  \,\V({\mathbf z},{\mathbf r},x)^{5/2}dx
\\
&=&E^{{\rm TF}}({\mathbf z},{\mathbf r})+D(\rho^{\rm TF}).
\nonumber
\eea
In Section~4 we shall make the semi-classical approximation more precise.

\section{New coherent states}

Coherent states provide a natural semi-classical description of quantum mechanics.
We shall denote these states by $|u,q\rangle$, where $(u,q)\in \R^{2n}$ is a point in
phase-space. Their wave-function is given by
\begin{equation}\label{old coherent states}\langle x|u,q\rangle = (\pi h)^{-n/4} e^{-(x-u)^2/2h} 
 e^{iqx/h}.
\end{equation}
Let $\Pi_{u,q} = |u,q\rangle\langle u,q|$ be the projection onto the coherent state
$|u,q\rangle$, then they satisfy the completeness condition (in the sense of quadratic forms)
\begin{equation}\label{resolution}
   \int \Pi_{u,q}\, \frac{dudq}{(2\pi h)^n} = {\bf 1}.
\end{equation}
As functions on phase-space the coherent states are localized on a scale of the order of
$h$. We want to broaden this by defining the operator
\begin{equation} \label{new coherent states}
     {\mathcal G}_{u,q}:= \int w(u-u',q-q') \, \Pi_{u',q'}\, du' dq'
\end{equation}
with
$$w(u,q) = \left(\frac{a}{\pi(1-ha)}\right)^n  e^{-a/(1-ha)\, (u^2+q^2)}.
$$
The new scale is $1/a>h$, which becomes clearer when we look at its kernel,
\begin{equation}{\mathcal G}_{u,q}(x,y) = (\pi h)^{-n/2} e^{-a\left(\frac{x+y}{2}-u\right)^2
      +iq(x-y)/h -\frac{1}{4h^2a}(x-y)^2} . \label{eq:Gkernel}
  \end{equation}
For simplicity, we have chosen a Gaussian weight, $w$, in
the definition of ${\mathcal G}_{u,q}$. We shall use the operators
${\mathcal  G}_{u,q}$ as our new coherent states.
\footnote{Sometimes (e.g.~see \cite{Lieb1}) it is useful to consider other coherent states
where the Gaussian
function in (\ref{old coherent states}) is replaced by a general $L^2$ function.
Similarly, one could use them to define corresponding generalized coherent states
but from a computational point of view the above choice is the simplest.

These new coherent states should not be confused with the quantum coherent operators
introduced by Lieb and Solovej in \cite{Lieb-Solovej} in order to compare two quantum
systems.}
Note that if we let $a\to 1/h$ then ${\mathcal G}_{u,q}$ converges to $\Pi_{u,q}$.
A straightforward calculation gives the following result.
\begin{lemma}[Completeness of new coherent states]
These new coherent operators satisfy
\begin{equation}\label{eq:resolution}
  \int {\mathcal G}_{u,q}^2 \,\frac{dudq}{(2\pi h)^n} = {\bf 1} .
\end{equation}
\end{lemma}
This resolution of the identity provides us with a representation of Schr\"odinger operators
as phase-space integrals. This will be useful when we
prove a lower bound on the sum of the negative eigenvalues of Schr\"odinger operators.

\begin{theorem}[Coherent states representation]\label{thm:coherentrepresentation}
Consider functions $F$ and $V$ in $C^3(\R^n)$, for which all second and third
derivatives are bounded. Let $\s(u,q)=F(q)+V(u)$, then we have for
$a<1/h$ and $b=2a/(1+h^2a^2)$ the representation
\begin{equation} \label{repr:coherentrepresentation}
F(-ih\nabla)+V(\x) = \int \,{\mathcal G}_{u,q}\widehat H_{u,q}
     {\mathcal G}_{u,q}\frac{du dq}{(2\pi h)^n} + {\mathbf E}
\end{equation}
as quadratic forms on $C^\infty_0(\R^n)$ with the operator-valued symbol
\begin{equation}\label{eq:coherentrepresentation}
\widehat H_{u,q}= \s(u,q)+\frac{1}{4b}\D \s(u,q)
+ \p_u \s(u,q)({\x}-u) + \p_q \s(u,q)(-ih\nabla -q).
\end{equation}
The error term, ${\mathbf E}$, is a bounded operator with operator norm
\begin{equation}\label{error}
\|\mathbf E\|\leq Cb^{-3/2}\sum_{|\alpha|=3}\|\partial^\alpha
\s\|_\infty+Ch^2b\sum_{|\alpha|=2}\|\partial^\alpha \s\|_\infty .
\end{equation}
\end{theorem}

Starting with the identity (\ref{eq:resolution}), the representation
of Schr\"odinger operators as in (\ref{repr:coherentrepresentation}) arises by
splitting the product ${\mathcal G}_{u,q}^2$ apart while sandwiching the symbol
$\hat{H}_{u,q}$. This operator-valued symbol can be thought of as the first order
Taylor expansion of the classical symbol $\s(u,q)$ at $(\hat{x},-ih\nabla)$.
Clearly, one could consider higher order expansions but this in not needed here.
Also notice that as $a\downarrow 1/h$ the linear term in
(\ref{eq:coherentrepresentation}) does not contribute in (\ref{repr:coherentrepresentation})
and one gets the familiar classical approximation $\s(u,q) + \frac{h}{4} \D\s(u,q)$.
The representation (\ref{repr:coherentrepresentation}) is symmetric in space and momentum
due to the symmetric Gaussian weights in the definition of ${\mathcal G}_{u,q}$.

One major advantage of coherent states 
is that a positive (upper) symbol leads to a positive operator.
This is important when writing down explicit variational states and
brings us to consider more generally operators of the form
\begin{equation}\label{eq:formf}
  \int {\mathcal G}_{u,q}\,f({\widehat A}_{u,q})\,{\mathcal G}_{u,q}\,dudq.
\end{equation}
Here, ${\widehat A}_{u,q}=B_0(u,q)+B_1(u,q)\cdot {\x}-ih B_2(u,q)\cdot\nabla$
is a Hermitian operator which is linear in ${\x}$ and $-ih\nabla$, and
$f:\R\to\R$ is any polynomially bounded real function.
We shall denote by $A_{u,q}$ the linear function
$A_{u,q}(v,p)=B_0(u,q)+B_1(u,q)\cdot v+B_2(u,q)\cdot p$.
When $A_{u,q}(v,p)$ is independent of $(v,p)$, i.e., if $B_1=B_2=0$
and if $a\to h^{-1}$ we recover the usual coherent states representation of an operator.
Thus on the one hand, we do not use as sharp a phase-space localization as the
one-dimensional coherent state projection since $a<1/h$, but
on the other hand, we use a better approximation than if $A_{u,q}$ were just a constant.

\section{Proof of semi-classical estimates}

\subsection{Regular potentials}

The key application of coherent states will be a proof of a semi-classical
expansion of the sum of negative eigenvalues of (localized) Schr\"odinger operators.
We shall restrict ourselves to localization functions supported in balls. Recall that we use
the convention, $x_-=(x)_- = \min\{x,0\}$.

\begin{theorem}[Local semi-classics] \label{corollary}
Let $n\geq 3$, $\phi\in C^{n+4}_0(\R^n)$ be supported in a ball
$B_\ell$ of radius $\ell>0$. Let $V\in C^3(\bar{B}_\ell)$ be a real potential.
Let $H=-h^2\D +V$, $h>0$ and $\sigma(u,q) = q^2 + V(u)$. Then for all
$h>0$ and $f>0$ we have
\begin{equation} \label{eq:phiHphilf}
  \left|\Tr[\phi H\phi]_- - (2\pi h)^{-n}\int \phi(u)^2 \sigma(u,q)_-\, du dq \right|
  \le C h^{-n+6/5} f^{n+4/5}\ell^{n-6/5},
\end{equation}
where the constant $C$ depends only on the dimension $n$,
\begin{equation}\label{eq:phivdependence}
  \sup_{|\a|\le n+4}\|\ell^{|\a|}\p^\a\phi\|_{\infty},\quad\hbox{ and }\quad
  \sup_{|\a|\le3}\|f^{-2}\ell^{|\a|}\p^\a V\|_{\infty}.
\end{equation}
Moreover, there exists a density matrix $\gamma$ such that
\begin{equation}
\Tr[\phi H\phi\gamma]\leq (2\pi h)^{-n}\int \phi(u)^2 \sigma(u,q)_-\, du dq
  +C h^{-n+6/5} f^{n+4/5}\ell^{n-6/5}\label{eq:gammaproplf} ,
\end{equation}
and such that its density  $\rho_\gamma(x)$ satisfies
\begin{equation}\label{eq:rhogammaproplf1}
\left|\rho_\gamma(x)-(2\pi h)^{-n}\omega_n
  \left|V(x)_-\right|^{n/2}\right|\leq Ch^{-n+9/10}f^{n-9/10}\ell^{-9/10},
\end{equation}
for (almost) all $x\in B_\ell$, and
\begin{equation}\label{eq:rhogammaproplf2}
\left|\int\phi(x)^2\rho_\gamma(x)dx-(2\pi h)^{-n}\omega_n\int\phi(x)^2
  \left|V(x)_-\right|^{n/2}dx\right|\leq Ch^{-n+6/5}f^{n-6/5}\ell^{n-6/5} .
\end{equation}
The constants $C>0$ in the above estimates again depend only on the dimension $n$,
the parameters in (\ref{eq:phivdependence}), and the volume of the unit ball
in $\R^n$, $\omega_n$.
\end{theorem}

As mentioned in the Introduction, any power ${\mathcal O}(h^{-n+1+\e})$ with $\e>0$ is sufficient
to prove the Scott correction in the main theorem (\ref{main theorem}). The power $\frac{6}{5}$ 
comes from optimizing the error bound in (\ref{error}) by choosing $b=h^{-\frac{4}{5}}$.

\begin{proof}[Sketch of proof] By a simple scaling argument we may restrict ourselves to the 
unit ball setting $\ell=1$ and the case $f=1$. We start with a sketch of the lower bound. We may 
assume that $h$ is sufficiently small. Using the representation (\ref{repr:coherentrepresentation}) 
we have that
\begin{eqnarray}
  {\Tr}[\phi H\phi]_-&\geq& {\Tr}\left[\int \phi
    \,{\mathcal G}_{u,q}\widehat{H}_{u,q}
    {\mathcal G}_{u,q}\phi\frac{du dq}{(2\pi h)^n}\right]_-\nonumber \\
  &&+ {\Tr}\left[\phi\left(-\epsilon h^2\Delta
      -C(b^{-3/2}+h^2b)\right)
    \phi\right]_- \label{eq:LTerror} .
\end{eqnarray}
Here, $0<\epsilon<1/2$, and
$$
\widehat H_{u,q}=\widetilde{\s}(u,q)+\frac{1}{4b}\D \widetilde{\s}(u,q)
+ \p_u\widetilde{\s}(u,q)({\x}-u) + \p_q \widetilde{\s}(u,q)(-ih\nabla -q)
$$
with $\widetilde{\s}(u,q)=(1-\epsilon)q^2+V(u)$. Utilizing the variational principle
for the first trace and the LT inequality for the second one we obtain the bound
$$ (2\pi h)^n
  {\Tr}[\phi H\phi]_-\geq \int {\Tr}\left[\phi
    \,{\mathcal G}_{u,q}\left[\widehat H_{u,q}\right]_-
    {\mathcal G}_{u,q}\phi\right] {du dq}
  - C \e^{-n/2}(b^{-3/2}+h^2b)^{1+n/2}.
$$
We shall eventually choose $\e=\frac{1}{4}(b^{-3/2}+h^2b)$. Since $\widehat{H}_{u,q}$ is a linear 
combination of $\x$ and $\nabla$ this operator can be easily rotated
into the momentum operator alone (up to some constant term). Then, we conveniently have an
expression for the negative part of $\widehat{H}_{u,q}$, and a fortiori, the trace becomes a
Gaussian-like integral which we have to estimate. In this integral, the linear function
$$
H_{u,q}(v,p)=\widetilde{\s}(u,q)+\frac{1}{4b}\Delta\widetilde{\s}(u,q)
             +\partial_u\widetilde{\s}(u,q)(v-u) + \partial_q \widetilde{\s}(u,q)(p-q)
$$
replaces the operator kernel of $\widehat H_{u,q}$; notice that
$(\x,-ih\nabla)$ is simply substituted by $(v,p)$.

We can show that if we consider the $u$-integration over $u$ outside the ball
$B_2$ of radius 2 then this is bounded below by $-Cb^{-3/2}$. On the other
hand, the integration over $B_2$ can be estimated from below by
$$\displaystyle\int_{u\in B_2}\phi^2\left(v+h^2ab(u-v)\right) G_b(u-v)G_b(q-p)
  \left[H_{u,q}(v,p)\right]_- {du dq} dp dv ,
$$
with $G_b(v)=(b/\pi)^{n/2}\exp(-bv^2)$. Since we are looking for a lower bound we may as
well extend the last integral to $\R^n$. Notice that we may now perform the $p$-integration
and obtain
\begin{eqnarray}
  (2\pi h)^n {\Tr}[\phi H\phi]_-&\geq&-\frac{2\omega_n}{n+2}
  (1-\varepsilon)^{-\frac{n}{2}}
  \int\phi^2\left(v+h^2abu\right) G_b(u)G_b(q)\nonumber
\\&&\quad\times
  \biggl|\left[V(v)+\widetilde{\xi}_v(u,q)-
    C|u|(b^{-1}+|u|^2)\right]_-\biggr|^{\frac{n}{2}+1} {du dq} dv
  \nonumber\\&&-C (b^{-3/2}+h^2b),\label{eq:lowerexpansion}
\end{eqnarray}
where we have introduced the function
$$
\widetilde{\xi}_v(u,q)
=\frac{1}{4b}
  \Delta\widetilde{\s}(v,0) -(1-\epsilon)q^2
  -\frac{1}{2}\sum_{ij}\partial_i\partial_j
  V(v)u_iu_j.
$$
By expanding we find that
\begin{eqnarray*}\lefteqn{
  \biggl|\left[V(v)+\widetilde{\xi}_v(u,q)-
    C|u|(b^{-1}+|u|^2)\right]_-\biggr|^{\frac{n}{2}+1}
}\\  
  &\leq& |V(v)|_-^{\frac{n}{2}+1}
  +\left(\frac{n}{2}+1\right)|V(v)_-|^{\frac{n}{2}}\widetilde{\xi}_v(u,q)\\
  &&
  +C\left(\left|\widetilde{\xi}_v(u,q)\right|+
    C|u|(b^{-1}+|u|^2\right)^2\\&&+C|u|(b^{-1}+|u|^2).
\end{eqnarray*}
We have used that since $n\geq 3$, the function $x\mapsto|x_-|^{\frac{n}{2}+1}$ is $C^2(\R^n)$.
Hence,
\begin{eqnarray*}
  (2\pi h)^n {\Tr}[\phi H\phi]_-&\geq&-\frac{2\omega_n}{n+2}
  (1-\varepsilon)^{-\frac{n}{2}}
  \int\phi^2\left(v+h^2abu\right) G_b(u)G_b(q)\\&&
  \times\,\left(|V(v)_-|^{\frac{n}{2}+1}
  + \left(\frac{n}{2}+1\right)|V(v)_-|^{\frac{n}{2}}\widetilde{\xi}_v(u,q)
  \right) {du dq} dv
\\
&&-C (b^{-3/2}+h^2b).
\end{eqnarray*}
We now expand $\phi^2$ and use the crucial identities for Gaussian integrals,
$$
 \int \widetilde{\xi}_v(u,q)G_b(u)G_b(q)dudq=0\quad
 \hbox{ and } \int u\,G_b(u)du=0 .
$$
We arrive at the lower bound,
\begin{eqnarray*}
  (2\pi h)^{n}{\Tr}[\phi H\phi]_-&\geq&-\frac{2\omega_n}{n+2}
  (1-\varepsilon)^{-\frac{n}{2}}
  \int\phi(v)^2|V(v)|_-^{\frac{n}{2}+1}dv-C(b^{-3/2}+h^2b)\\
  &=&(1-\varepsilon)^{-\frac{n}{2}}
  \int\phi(v)^2\s(v,p)_-dvdp-C(b^{-3/2}+h^2b).
\end{eqnarray*}
Finally, we choose $a=h^{-4/5}$ and $\epsilon=b^{-3/2}$.

Now we come to the upper bound. We shall show here only the construction of the density matrix
$\c$. Let $\chi=\chi_{(-\infty,0]}$ be the characteristic function of $(-\infty,0]$ and
\beax\lefteqn{\hat{h}_{u,q}}
\\
&&=\left\{\begin{array}{cl}\s(u,q)+ \frac{1}{4b}\Delta\s(u,q) + \p_u\s(u,q)({\x}-u) +
   \p_q {\s}(u,q)(-ih\nabla -q)&\!\!\!\!\hbox{if } u\in B_2\\0&\!\!\!\!\hbox{if } u\not\in B_2
\end{array}\right.
\eeax
We then define
\begin{equation} \label{trial density}
   \c = \int {\mathcal G}_{u,q}\,\chi\big[\hat{h}_{u,q} \big]\,
             {\mathcal G}_{u,q}\,\frac{dudq}{(2\pi h)^n} .
\end{equation}
Since ${\bf 0}\le\chi\big[\hat{h}_{u,q}\big]\le\bf1$ it is obvious that ${\bf 0}
\le\c\le{\bf 1}$. The arguments showing that
$${\Tr}[\c\phi H\phi] \le (2\pi h)^{-n}\int \,\phi^2(u) \s(u,q)_- du dq
  + Ch^{-n+6/5}
$$
are then very similar to the above calculations for the lower bound, see \cite{SS}.
\end{proof}

\subsection{Thomas-Fermi potential}

In this Section we shall sketch the proof of the Scott correction for the TF potential.

\begin{theorem}[Scott corrected semi-classics]\label{TF}
For all $h>0$ and all $r_1,\ldots,r_M$ $\in\R^3$ with $\min_{k\ne m}|r_m-r_k|>r_0>0$
we have
\bea\label{eq:main1}
\lefteqn{\left|\Tr[-h^2\D - V^{\rm TF}]_- - (2\pi h)^{-3} \int (p^2 - V^{\rm TF}(u))_-
 \,du dp - \frac{1}{8h^2} \sum_{k=1}^M z_k^2\right|
}
\nonumber\\
&&\phantom{\Tr[-h^2\D - V^{\rm TF}]_- - (2\pi h)^{-3} \int (p^2 - V^{\rm TF}(u))_-
  \sum_{k=1}^M z_k^2}
  \le C h^{-2+\frac{1}{10}},
\eea
where $C>0$ depends only on $z_1,\ldots,z_M$, $M$, and $r_0$.
Moreover, we can find a density matrix $\gamma$ such that
\begin{equation}\label{eq:maingamma1}
\Tr \left[(-h^2\Delta - V^{\rm TF})\gamma\right]
\leq \Tr \left[-h^2\Delta - V^{\rm TF}\right]_-+C h^{-2+1/10},
\end{equation}
and such that
\begin{equation}\label{eq:maingamma2}
D\left(\rho_\gamma-\frac{1}{6\pi^2h^3}(V^{\rm TF})^{3/2}\right)\leq Ch^{-5+4/5} ,
\end{equation}
and
\begin{equation}\label{eq:maingamma3}
    \int \rho_\gamma\leq \frac{1}{6\pi^2h^3}\int V^{\rm
      TF}(x)^{3/2}dx+C h^{-2+1/5},
  \end{equation}
  with $C$ depending on the same parameters as before.
\end{theorem}

\begin{proof}[Sketch of proof] From Theorem \ref{thm:tfestimate} we know that the TF potential 
has an inverse fourth power law decay far from the nuclei. Thus, a region outside some ball of 
radius $R$ (which scales with $h$) should contribute little to the sum of negative energies. For 
this purpose we introduce a first partition of unity. So let us choose
\begin{equation}\label{eq:Rchoice}
  R=h^{-1/2} ,
\end{equation}
and consider functions $\Phi_\pm\in C^\infty(\R^n)$ such that
\begin{enumerate}
\item $\Phi_-^2+\Phi_+^2=1$,
\item $\Phi_-(x)=1$ if $d(x)<R$ and $\Phi_-(x)=0$ if $d(x)>2R$.
\end{enumerate}
Denote ${\mathcal I}=(\nabla\Phi_-)^2+(\nabla\Phi_+)^2$.
Then ${\mathcal I}$ is supported on a set whose volume is bounded by $CR^3$
(where as before $C$ depends on $M$) and $\|{\mathcal I}\|_\infty\leq CR^{-2}$.
Using the standard IMS localization formula and then the LT inequality
we find that
\beax\lefteqn{
  \Tr[-h^2\Delta-V^{\rm TF}]_-}
\\
&\ge&\Tr[\Phi_-(-h^2\Delta-V^{\rm TF}-h^2{\mathcal I})\Phi_-]_-
  +\Tr[\Phi_+(-h^2\Delta-V^{\rm TF}-h^2{\mathcal I})\Phi_+]_-
\\
&\ge&\Tr[\Phi_-(-h^2\Delta-V^{\rm TF}-h^2{\mathcal I})\Phi_-]_- -C (h^2R^{-2} + h^{-3}R^{-7})  .
\eeax
With the chosen $R$ the last term is of the order $h^{-1/2}$.

On the support of $\Phi_-$ we want to use the hydrogenic approximation of the TF potential
close (of the order of $h$) to the nuclei, and on the rest the semi-classical estimates from 
the previous Section. Let us introduce the function
\begin{equation}\label{eq:ldefinition}
  \ell(x)=\mfr{1}{2}\Bigl(1+\sum_{k=1}^M(|x-r_k|^2+ h^2)^{-1/2}\Bigr)^{-1} .
\end{equation}
Note that $\ell$ is a smooth function with
$$
0<\ell(x)<1,\quad\hbox{and}\quad
\|\nabla\ell(x)\|_\infty<1.
$$
Now, we fix some localization function $\phi\in C_0^\infty(\R^3)$ with support
in the unit ball $\{|x|<1\}$ and such that $\int\phi(x)^2dx=1$. It is not difficult
(cf Theorem~22, \cite{SS}) to find a corresponding
family of functions $\phi_u\in C_0^\infty(\R^3)$, $u\in\R^3$,
where $\phi_u$ is supported in the ball $\{|x-u|<\ell(u)\}$, with the
properties that
\begin{equation}\label{eq:phiuprop}
  \int\phi_u(x)^2\ell(u)^{-3}du=1\quad\hbox{and}\quad
  \|\partial^\alpha\phi_u\|_\infty\leq C\ell(u)^{-|\alpha|},
\end{equation}
for all multi-indices $\alpha$, where $C>0$ depends only on $\alpha$ and $\phi$.

One can show from (\ref{eq:tfdf}) in
Theorem~\ref{thm:tfestimate} that for all $u\in\R^n$ with $d(u)>2h$,
\begin{equation}\label{eq:tflf}
  \sup_{|x-u|<\ell(u)}|\p^\alpha V^{\rm TF}(x)|\leq Cf(u)^2\ell(u)^{-|\alpha|},
\end{equation}
where $C>0$ depends only on $\alpha$, $z_1,\ldots,z_M$, and $M$. This is the requirement
for the semi-classical estimates from Theorem \ref{corollary} to apply with $\ell(u) \to \ell, 
f(u)\to f$.

Another application of the IMS formula shows that
\begin{eqnarray}
  \lefteqn{\hspace*{.7cm}\Tr[-h^2\Delta-V^{\rm TF}]_-}&&\label{eq:tfphilow}\\\nonumber
  &\geq&\int_{d(u)<2R+1}\Tr[\phi_u\left(-h^2\Delta-V^{\rm TF}-Ch^2\ell(u)^{-2}\right)\phi_u]_-
  \ell(u)^{-3}du - C h^{-1/2} .\nonumber
\end{eqnarray}
By similar arguments we get corresponding estimates for the
hydrogenic operators replacing $V^{\rm TF}$ by $\frac{z_k}{|x-r_k|}-1$ in the above estimates.
In particular, if we choose $h$ so small that $R>\max_k\{z_k\}$ then on the support of 
$\Phi_+$ we have $-z_k|x-r_k|^{-1}+1\geq0$. Thus we have
\begin{eqnarray}
  \lefteqn{\hspace*{.4cm}\Tr\Bigl[-h^2\Delta-\frac{z_k}{|\hat{x}-r_k|}+1\Bigr]_-}&&
  \label{eq:hydphilow}\\
  &\geq&\int_{d(u)<2R+1}\Tr\Bigl[\phi_u\Bigl(-h^2\Delta-\frac{z_k}{|\hat{x}-r_k|}+1
  -Ch^2\ell(u)^{-2}\Bigr)\phi_u\Bigr]_- \ell(u)^{-3}du\nonumber\\&&-Ch^{2} R^{-2}.\nonumber
\end{eqnarray}
We arrive at analoguous upper bounds if we utilize the density matrix
$$
\gamma=\int_{d(u)<2R+1}\phi_u\chi\left(\phi_u(-h^2\Delta-V^{\rm TF})\phi_u\right)\phi_u
\ell(u)^{-3}du
$$
as a trial operator. I.e.,
\begin{eqnarray}
  \Tr[-h^2\Delta-V^{\rm TF}]_-&\leq& \Tr[(-h^2\Delta-V^{\rm TF})\gamma]\nonumber\\
  &=&\int\limits_{d(u)<2R+1}\Tr[\phi_u\left(-h^2\Delta-V^{\rm
      TF}\right)\phi_u]_-  \ell(u)^{-3}du.\label{eq:tfphiup}
\end{eqnarray}
Similarly, 
\begin{eqnarray}\lefteqn{\hspace*{.0cm}\label{eq:hydphiup}
  \Tr\Bigl[-h^2\Delta-\frac{z_k}{|\hat{x}-r_k|}+1\Bigr]_-}
  \\
  &\leq&
  \int\limits_{d(u)<2R+1}\!\!\!\!\Tr\Bigl[\phi_u\Bigl(-h^2\Delta-\frac{z_k}{|\hat{x}-r_k|}+1\Bigr)\phi_u\Bigr]_-  
  \ell(u)^{-3}du\nonumber.
\end{eqnarray}
We now introduce the quantities
\begin{eqnarray*}
  D_+(u)&:=&\Tr[\phi_u\left(-h^2\Delta-V^{\rm
      TF}-Ch^2\ell(u)^{-2}\right)\phi_u]_-\nonumber\\&&-
  \sum_{k=1}^M\Tr\Bigl[\phi_u\Bigl(-h^2\Delta-\frac{z_k}{|\x-r_k|}+1\Bigr)\phi_u\Bigr]_-
   \\
  D_-(u)&:=&\sum_{k=1}^M\Tr\Bigl[\phi_u\Bigl(-h^2\Delta
  -\frac{z_k}{|\x-r_k|}+1-Ch^2\ell(u)^{-2}\Bigr)\phi_u\Bigr]_-
  \nonumber\\&&-
  \Tr[\phi_u(-h^2\Delta-V^{\rm
    TF})\phi_u]_- ,\\
\noalign{and}
D_{\rm SC}(u)&:=&(2\pi h)^{-3}\int\phi_u(x)^2(p^2-\V(x))_-dpdx\nonumber\\&&-
(2\pi h)^{-3}\sum_{k=1}^M\int\phi_u(x)^2\Bigl(p^2-\frac{z_k}{|x-r_k|}+1\Bigr)_-dpdx
\end{eqnarray*}
Then, from (\ref{eq:tfphilow}), and (\ref{eq:tfphiup}) we have
\begin{eqnarray}
  \lefteqn{\Tr[-h^2\Delta-\V]_- - \sum_{k=1}^M\Tr\Bigl[-h^2\Delta-\frac{z_k}{|\x-r_k|}+1\Bigr]_-
}
\\
 &\geq&\int_{d(u)<2R+1} D_+(u)\ell(u)^{-3}du -C h^{-1/2}\label{eq:Dintegral} \nonumber,
\end{eqnarray}
and from (\ref{eq:phiuprop}) we get
\begin{eqnarray}
  (2\pi h)^{-3}\int(p^2-\V(x))_-dpdx-
    (2\pi h)^{-3}\sum_{k=1}^M\int\Bigl(p^2-\frac{z_k}{|x-r_k|}+1\Bigr)_-dpdx
  \nonumber\\=
  \int D_{\rm SC}(u)\ell(u)^{-3}du.\label{eq:Dscintegral}
\end{eqnarray}
Next, we compute explicitly both the quantum and the semi-classical
energies for the Coulomb potential, namely
$$
\Tr\Bigl[-h^2\Delta-\frac{z_k}{|\x-r_k|}+1\Bigr]_-
=\sum_{1\leq n\leq z_k/(2h)}\left(-\frac{z_k^2}{4h^2}+n^2\right)
=-\frac{z_k^3}{12 h^3}+\frac{z_k^2}{8h^2}+{\mathcal O}(h^{-1}) ,
$$
and
$$
 (2\pi h)^{-3}\int \Bigl(p^2-\frac{z_k}{|u-r_k|}+1\Bigr)_-dudp
 =-\frac{z_k^3}{12 h^3}.
$$
The first statement of the theorem is thus proven once we establish lower bounds on
$D_+(u)-D_{\rm SC}(u)$ and $D_-(u)+D_{\rm SC}(u)$.
Here, we have to distinguish between the region $d(u)<2h$ and the semi-classical
region, $2h<d(u)<2R+1$.

In the region close to the nuclei, $d(u)<2h$, we use the estimate (\ref{eq:westimate})
on the potential, $W_k(x)=V^{{\rm TF}}(x)-z_k|x-r_k|^{-1}$. Then, all bounds on $D_\pm(u)$
are obtained via the LT inequality.

Secondly, let $2h<d(u)<2R+1$. On the ball $\{x\ |\ |x-u|<\ell(u)\}$, the TF potential satisfies the
estimate (\ref{eq:tflf}) and $\phi_u$ satisfies (\ref{eq:phiuprop}). Hence, we may use
Theorem~\ref{corollary}.  A similar semi-classical estimate holds for the Coulomb
potential, $-\frac{z_k}{|x-r_k|}+1$.

The density matrix which we choose for (\ref{eq:maingamma1}--\ref{eq:maingamma3}) to hold
is constructed as follows. If $2h<d(u)$ it follows from (\ref{eq:phiuprop}),
(\ref{eq:tflf}), and Theorem~\ref{corollary}
that we may choose  $\gamma_u$ such that (\ref{eq:gammaproplf}),
(\ref{eq:rhogammaproplf1}), and  (\ref{eq:rhogammaproplf2}) hold
when $V=V^{\rm TF}$, $\phi=\phi_u$, $\ell=\ell(u)$ and $f=f(u)$.

If $d(u)\leq 2h$ we simply choose
$$
  \gamma_u=
  \chi\left[\phi_u\left(-h^2\Delta-V^{\rm TF}\right)\phi_u\right],
$$
where $\chi$ is again the characteristic function of the interval
$(-\infty,0]$. I.e., $\gamma_u$ is the projection onto the
non-positive spectrum of $\phi_u\left(-h^2\Delta-V^{\rm TF}\right)\phi_u$. Here we
are considering $\phi_u$ as a multiplication operator.

Finally, we set
\begin{equation}\label{eq:gammachoice}
  \gamma=\int_{d(u)<R} \phi_u\gamma_u\phi_u\ell(u)^{-3}du .
\end{equation}
By the properties (\ref{eq:phiuprop}), $\c$ is a density matrix.

\end{proof}

\section{Proof of main theorem}

The proof of the main theorem on the molecular ground state energy is a rather standard
application of the results presented in the previous Sections. We prove lower and upper
bounds.
\begin{proof}[Proof of Theorem \ref{main theorem}]
The starting point for a lower bound is the Lieb-Oxford inequality (\ref{Lieb-Oxford})
from which we conclude that if $\psi$ is a $Z$-particle ($N=Z$) then
$$
\langle \psi, H(Z,R)\psi\rangle
\ge \sum_{i=1}^Z \left\langle\psi,\big[-\mfr{1}{2}\D_i -
  V(Z,R,x_i)\big]
  \psi\right\rangle
+ D(\rho_\psi) - C\int \rho_\psi^{4/3}.
$$
In order to bound the last term we use the
many-body version of the LT inequality (\ref{eq:LTmbcase}).
For all $0<\e<1/2$ we have
\begin{eqnarray*}
  \Big\langle\psi,\e\sum_{i=1}^Z-\mfr{1}{2}\D_i \psi\Big\rangle
  - C\int \rho_\psi^{4/3}
  &\ge&-\e^{-1} C \int \rho_\psi =-C\e^{-1}Z.
\end{eqnarray*}
Here we have used H\"older's inequality for the $\rho^{4/3}$ integral
and the assumption that $\psi$ is a $Z$-particle state.
Thus
\begin{eqnarray*}
\lefteqn{\langle \psi, H(Z,R)\psi\rangle}
\\
&\ge&\left\langle \psi,\sum_{i=1}^Z (-(1-\e)\mfr{1}{2}\D_i - V(Z,R,x_i))
     \psi\right\rangle + D(\rho_\psi) - C\e^{-1}Z
\\
&\geq&2\,\Tr\big[-\mfr{1}{2}(1-\e)\D - \V(Z,R,\cdot)\big]_-
      - D(\rho^{\rm TF}(Z,R,\cdot))- C\e^{-1}Z.
\end{eqnarray*}
Here we have applied (\ref{eq:tfpotgeneral}), the fact that the Coulomb
kernel is positive definite such that $D(\rho-\rho^{\rm TF})\geq0$, and the
fermionic property of the wave function.

If we now use the scaling property (\ref{scaling}) we find that
$$
\Tr\big[-\mfr{1}{2}(1-\e)\D - \V(Z,R,\cdot)\big]_-
=|Z|^{4/3}\Tr\big[-\mfr{1}{2}(1-\e)|Z|^{-2/3}\D -
\V({\bf z},{\bf r},\cdot)\big]_-,
$$
where ${\bf z}=(z_1,\ldots,z_M)$ and ${\bf r}=(r_1,\ldots,r_M)$.
Using now (\ref{eq:main1}) (with $h=\sqrt{\frac{1-\e}{2}}\cdot$ $|Z|^{-1/3}$)
and (\ref{eq:sc=tf}) we see that
\beax\lefteqn{
2\Tr\big[-\mfr{1}{2}(1-\e)\D - \V(Z,R,\cdot)\big]_- }
\\
&=&(1-\e)^{-3/2}|Z|^{7/3}\left(E^{\rm TF}({\bf z},{\bf r})+D(\rho^{\rm TF}({\bf
      z},{\bf r},\cdot)\right)
  +(1-\e)^{-1}\frac{|Z|^2}{2}\sum_{k=1}^Mz_k^2
\\
&&+O(|Z|^{2-1/30})\\
&=&(1-\e)^{-3/2}\left(E^{\rm TF}(Z,R)+D(\rho^{\rm TF}(Z,R,\cdot)\right)
  +\mfr{1}{2}(1-\e)^{-1}\sum_{k=1}^MZ_k^2
\\
&&+O(|Z|^{2-1/30}).
\end{eqnarray*}
We have here used the TF scaling $E^{\rm TF}(Z,R)=|Z|^{7/3}E^{\rm
  TF}({\bf z},{\bf r})$ and $D(\rho^{\rm
  TF}\left(Z,R,\cdot\right)$ $=|Z|^{7/3}D\left(\rho^{\rm TF}({\bf z},{\bf
    r},\cdot)\right)$. Choosing $\e=|Z|^{-2/3}$ completes the proof
of the lower bound.

The starting point for an upper bound is Lieb's variational principle,
Theorem~\ref{Lieb's Variational Principle}.
By a simple rescaling the variational principle states that for any
density matrix $\gamma$ on $L^2(\R^3)$ with $2\Tr\gamma\leq Z$ we have
$$
  E(Z,R)\leq|Z|^{4/3}\left(2\Tr\left[\left(-\mfr{1}{2}|Z|^{-2/3}\Delta
      -V({\bf z},{\bf r},x)\right)\gamma\right]
  +|Z|D(2|Z|^{-1}\rho_\gamma)\right).
$$
As for the lower bound we bring the TF-potential into play,
\begin{eqnarray}
  |Z|^{-4/3}E(Z,R)&\leq&2\Tr\left[\left(-\mfr{1}{2}|Z|^{-2/3}\Delta
      -V({\bf z},{\bf r},x)\right)\gamma\right]
  +|Z|D(2|Z|^{-1}\rho_\gamma)\nonumber\\
  &=&
  2\Tr\left[\left(-\mfr{1}{2}|Z|^{-2/3}\Delta
      -\V({\bf z},{\bf r},x)\right)\gamma\right]\nonumber\\&&
  +|Z|D\left(2|Z|^{-1}\rho_\gamma-\rho^{\rm TF}({\bf z},{\bf r},\cdot)\right)
  -|Z|D(\rho^{\rm TF}({\bf z},{\bf r},\cdot)).\label{eq:upper1}
\end{eqnarray}
We now choose a density matrix $\widetilde{\gamma}$
according to Theorem~\ref{TF} with $h=\sqrt{1/2}|Z|^{-1/3}$.
Note that with this choice of $h$ we have that
$$
(6\pi^2h^3)^{-1}\V({\bf z},{\bf r},x)^{3/2}=|Z|\rho^{\rm TF}({\bf
  z},{\bf r},x)/2.
$$
Since $\int\rho^{\rm TF}({\bf z},{\bf
  r},x)=\sum_{j=1}^Mz_j=1$ we see from (\ref{eq:maingamma3}) that
$$
2\Tr\widetilde\gamma\leq |Z|+C|Z|^{2/3-1/15}=|Z|(1+C|Z|^{-1/3-1/15}).
$$
Thus if we define $\gamma=(1+C|Z|^{-1/3-1/15})^{-1}\widetilde{\gamma}$
we see that the condition $2\Tr\gamma\leq |Z|$ is satisfied.

Using (\ref{eq:maingamma2}) we conclude that
$$
|Z|D\left(2|Z|^{-1}\rho_{\widetilde\gamma}-\rho^{\rm TF}({\bf z},{\bf
    r},\cdot)\right) \leq C|Z|^{2/3-4/15} ,
$$
and thus
\begin{eqnarray}
 |Z|D\left(2|Z|^{-1}\rho_\gamma-\rho^{\rm TF}({\bf z},{\bf
        r},\cdot)\right)\leq C |Z|^{2/3-4/15},\label{eq:upper2}
\end{eqnarray}
where we have used that $D\left(\rho^{\rm TF}({\bf z},{\bf
    r},\cdot)\right)\leq C$.

Finally, if we use (\ref{eq:main1}), (\ref{eq:maingamma1}), and
(\ref{eq:sc=tf}) we arrive at
\begin{eqnarray*}
  2\Tr\left[\left(-\mfr{1}{2}|Z|^{-2/3}\Delta
      -\V({\bf z},{\bf r},x)\right)\widetilde\gamma\right]&\leq&
  |Z|\left(E^{\rm TF}({\bf
      z},{\bf r})+D(\rho^{\rm TF}({\bf
      z},{\bf r},\cdot)\right)\\&&
  +\frac{|Z|^{2/3}}{2}\sum_{k=1}^Mz_k^2+O(|Z|^{2/3-1/30}).
\end{eqnarray*}
Since $E^{\rm TF}({\bf z},{\bf r})\leq C$ and
$D\left(\rho^{\rm TF}({\bf z},{\bf r},\cdot)\right)\leq C$ we see that the
same estimate holds for $\widetilde\gamma$ replaced by $\gamma$.
If we insert this estimate together with (\ref{eq:upper2}) into
(\ref{eq:upper1}) and use again that
$E^{\rm TF}(Z,R)=|Z|^{7/3}E^{\rm TF}({\bf z},{\bf r})$ we arrive at
the upper bound.
\end{proof}

\bibliographystyle{amsalpha}

\begin{thebibliography}{A}

\bibitem [B]{Bach} V.~Bach: \textit{A proof of {S}cott's conjecture for ions}, Rep.~Math.~Phys.,
    {\bf 28} 213--248 (1989)    

\bibitem [H]{Hughes} W.~Hughes: \textit{An atomic energy bound that gives Scott's correction},
    Adv.~Math., {\bf 79} 213--270 (1990)

\bibitem [IS]{Ivrii-Sigal} V.I.~Ivrii and I.M.~Sigal: \textit{Asymptotics of the ground state
    energies of large Coulomb systems}, Ann.~of Math. (2) {\textbf 138}, 243--335 (1993)

\bibitem [L1]{Lieb1} E.H.~Lieb: \textit{Thomas-Fermi theories and related theories of atoms and
     molecules}, Rev.~Mod.~Phys. {\textbf 53}, 603--641 (1981)

\bibitem [L2]{Lieb2} E.H.~Lieb: \textit{Variational principle for many-fermion systems},
     Phys.~Rev.~Lett, vol. {\textbf 46}, 457--459 (1981), or in
     \textit{The stability of matter: from atoms to stars}, Springer (1991)

\bibitem [L3]{Lieb3} E.H.~Lieb: \textit{A lower bound for Coulomb energies}, Phys.~Lett.
     {\textbf 70A}, 444--446 (1979)

\bibitem [L4]{Lieb:sob} E.H.~Lieb: \textit{Sharp constants in the Hardy-Littlewood-Sobolev and
     related inequalities}, Ann.~of Math., {\textbf 118} no. 2, 349--374, (1983)

\bibitem [LL]{Lieb-Loss} E.H.~Lieb and M.~Loss: \textit{Analysis}, Graduate studies in Mathematics,
     vol.~{\textbf 14} (2001)

\bibitem [LO]{Lieb-Oxford} E.H.~Lieb and S.~Oxford: \textit{An improved lower bound on the indirect
    Coulomb energy}, Int.~J.~Quant.~Chem. {\textbf 19}, 427--439 (1981)
    
\bibitem [LSi]{Lieb-Simon} E.H.~Lieb and B.~Simon: \textit{Thomas-Fermi theory of atoms, molecules
    and solids}, Adv.~in Math., {\textbf 23}, 22--116, (1977)

\bibitem[LSo]{Lieb-Solovej} E.H.~Lieb and J.P.~Solovej: {\em Quantum coherent operators: 
    A generalization of coherent states}, Lett.~Math.~Phys., {\bf 22}, 145--154 (1991)

\bibitem [M]{Matesanz} P.~Balodis Matesanz: \textit{A proof of Scott correction for Matter}, 
    preprint mp-arc/02-62 

\bibitem [MS]{Matesanz-Solovej} P.~Balodis Matesanz and J.P.~Solovej: \textit{On the asymptotic 
    exactness of Thomas-Fermi theory in the thermodynamic limit},  Ann.~Henri Poincare, {\bf 1}, 
    281--306 (2000)

\bibitem [LT]{Lieb-Thirring} E.H.~Lieb and W.E.~Thirring: \textit{Inequalities for the moments
    of the eigenvalues of the Schr\"odinger Hamiltonian and their relation to Sobolev
    inequalities,} in Studies in mathematical physics, (E.~Lieb, B.~Simon,
    and A.S.~Wightman, eds.), Princeton Univ. Press, Princeton, New Jersey,
    269--330 (1976)

\bibitem [SW]{Siedentop-Weikard} H.~Siedentop and R.~Weikard: \textit{On the leading energy 
     correction for the statistical model of an atom: interacting case}, 
     Comm.~Math.~Phys.~{\textbf 112}, 471--490 (1987), \textit{On the leading correction of 
     the Thomas-Fermi model: lower bound}, Invent.~Math. {\textbf 97}, 159--193 (1990), and 
     \textit{A new phase space localization technique with application to
     the sum of negative eigenvalues of {S}chr\"odinger operators}, Ann.~Sci.~\'Ecole 
     Norm. Sup. (4), vol.~{\bf 24}, no.~2, 215--225 (1991)

\bibitem [SS]{SS} J.P.~Solovej and W.L.~Spitzer: \textit{New coherent states approach to
     semiclassics which gives Scott's correction,} preprint mp-arc/02-357, or 
     math-ph/0208044 (2002)

\bibitem [T]{Thirring} W.~Thirring: \textit{A lower bound with the best possible
     constant for Coulomb Hamiltonians}, Commun.~Math.~Phys., {\textbf 79} no.~1, 1--7, (1981)

\end{thebibliography}

\end{document}